\documentclass[printer]{aa}
\usepackage{psfig}

\def\ltsima{$\; \buildrel < \over \sim \;$}
\def\lsim{\lower.5ex\hbox{\ltsima}}
\def\gtsima{$\; \buildrel > \over \sim \;$}
\def\gsim{\lower.5ex\hbox{\gtsima}}

\begin{document}

\title{Thermal components in the early X-ray afterglow of GRBs}
\titlerunning{Thermal X-ray afterglows}

\author{Davide Lazzati}
\authorrunning{D. Lazzati}
\institute{Institute of Astronomy, University of Cambridge, Madingley Road,
Cambridge CB3 0HA, England \\
e-mail: {\tt lazzati@ast.cam.ac.uk}}

\date{}

\abstract{
The possible presence of thermal components in the early X-ray
afterglows of $\gamma$-ray bursts is investigated. We discuss both the
presence of a thermal continuum and, in particular, of collisional
X-ray emission lines. We compute the predicted luminosity by a thin
plasma for a range of metallicities for the continuum and the
$K_\alpha$ lines of the elements Mg, Si, S, Ar, Ca and Fe.  We show
that light travel effects are dominant in the determination of the
thermal continuum and line luminosities, and derive the relevant
equations.  We conclude that thermal lines and continua are unlikely
to dominate the early afterglow of GRBs, unless the explosion site is
surrounded by a very massive and extremely clumped shell of
material. Such conditions are difficult to envisage in the close
environment of GRB progenitor, unless they are excited by some strong
precursor activity, like in the Supranova scenario.
}

\maketitle

\section{Introduction}

Almost all the $\gamma$-ray bursts (GRBs) detected so far are
associated to a transient X-ray afterglow (Lazzati et al. 2002a).  This
afterglow is supposed to be due to the early deceleration of the
fireball by the interstellar medium (Meszaros \& Rees 1997) and its
radiation produced by non-thermal synchrotron. Such an interpretation
is corroborated by spectral (van Paradijs et al. 2002) and
polarimetric (Covino et al. 1999) observations. 

More recently, several X-ray afterglows observed by Newton--XMM had a
spectrum that can be better fit by an optically thin thermal
bremsstrahlung model (Watson et al. 2002) rather than with an absorbed
power-law (even though the latter model cannot be unambiguously ruled
out on purely statistical grounds). In particular, several high
ionization emission lines were detected in the early afterglow of
GRB~011211 (Reeves et al. 2002, hereafter R02) which can be accounted
for, together with the observed continuum, by a moderately enriched
thermal plasma.

The possibility of a thermal origin of the X-ray lines detected in
several afterglows was first discussed in Lazzati et al. (1999; see
also Vietri et al. 1999 and Kumar \& Narayan 2002), who pointed
out how a 10 times solar plasma may produce a Fe $K_\alpha$ line with
the observed large luminosity and equivalent width. In this paper we
derive more rigorously the conditions for the environment and the
heating mechanism that must be satisfied in order to observe line and
continuum emission with the prescribed luminosities, equivalent widths
(EWs) and variability time scales.

\section{Continuum and line emission in thermal plasma}

\begin{figure*}
\centerline{\psfig{file=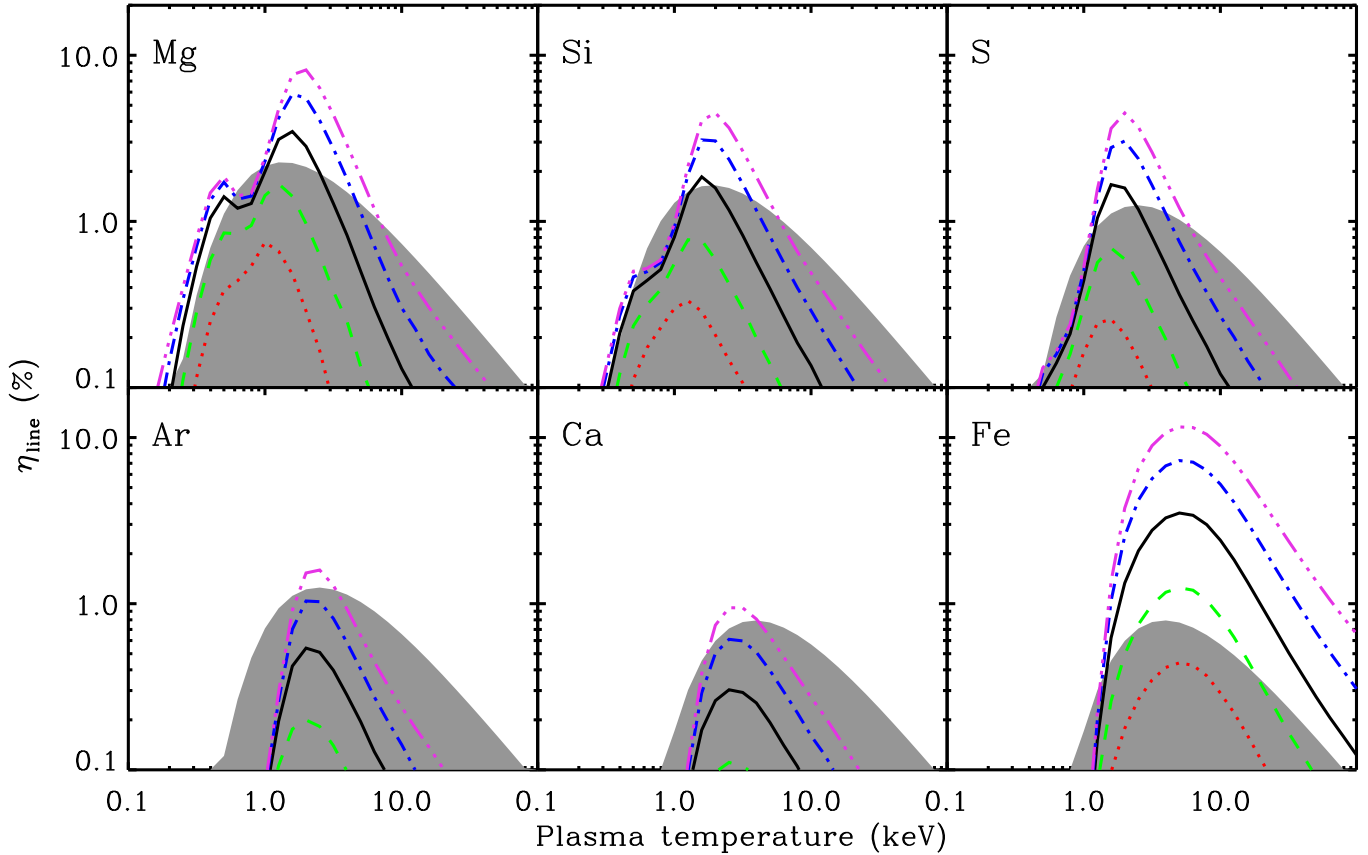,width=\textwidth}}
\caption{{Efficiency of $K_\alpha$ line emission for several 
elements as a function of the temperature in an optically thin
plasma. The parameter $\eta_{\rm{line}}$ is defined as the line
luminosity over the bolometric luminosity of the thermal plasma.  Each
panel shows 5 different lines. From top to bottom these refer to a
metallicity 10, 3, 1, 0.3 and 0.1 solar. The gray shading highlights
regions in the parameter space where an equivalent width lower than
100~eV is predicted.}
\label{fig:eta}}
\end{figure*}

To compute the continuum emission from a thermal optically thin plasma
we adopt the treatment and basic equations of Rybicki \& Lightman
(1979).  To compute the luminosities of $K_\alpha$ lines from a
thermal plasma, instead, we adopt the widely used code MEKAL (Mewe et
al. 1985; Liedahl et al. 1995) as implemented in XSPEC (Arnaud
1996). We concentrate in particular the elements Mg, Si, S, Ar, Ca and
Fe, for which emission lines have been detected in the afterglow of
GRBs.  In Fig.~\ref{fig:eta} we show the line production efficiency
for an optically thin thermal plasma as a function of the plasma
temperature and for several values of metallicity. We plot
$\eta_{\rm{line}}$, i.e.  the ratio of the $K_\alpha$ emission line
for the six elements above, irrespective of their ionization state,
over the total luminosity of the plasma. Gray shading highlights
regions in which the equivalent width (EW) of the lines is less than
100~eV, a robust lower limit to any emission feature detected in the
afterglow so far.

\begin{figure}
\psfig{file=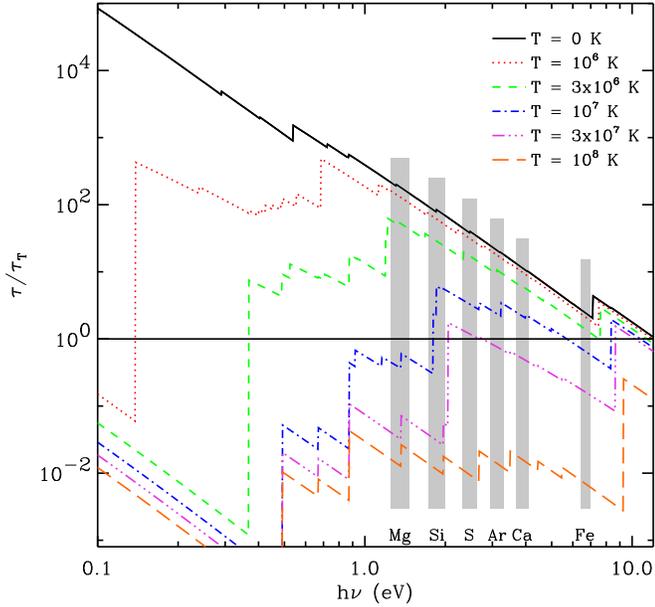,width=0.48\textwidth}
\caption{{Photoionization optical depth of a $\tau_T=1$ plasma in thermal 
equilibrium as a function of the temperature and frequency. The range
of frequencies of the considered $K_\alpha$ lines is shown by gray
shading.}
\label{fig:tau}}
\end{figure}

When using the MEKAL code to evaluate line emission luminosities, one
has to remember that the code does not include any radiation transfer,
since it assumes that the medium is optically thin to radiation. The
actual optical depth of a cloud of plasma at temperature $T$ depends
on the temperature and on the frequency of the radiation
considered. When X-ray continuum radiation is considered, the
optically thin approximation can be used up to column densities
$N_H\lsim\sigma_T^{-1}\sim1.5\times10^{24}$~cm$^{-2}$. If, however,
line emission is concerned, it must be taken into account that
intermediate-high $Z$ elements retain some electrons which may cause
the plasma to be optically thick due to photoionization. In
Fig.~\ref{fig:tau} we show the optical depth of a solar metallicity
plasma with $N_H=1.5\times10^{24}$~cm$^{-2}$ as a function of
frequency for a range temperatures between $10^6$ and $10^8$~K. The
opacity of a cold gas is also shown for comparison. In the region of
the considered emission lines, the plasma can be optically thick up to
temperatures of several keV. This will limit the maximum line
luminosity and EW: increasing the column density of the plasma will
have no effect on the line luminosity since line photons will be able
to escape freely only out of the optically thin surface layer of the
medium. Instead of properly introducing radiation transfer in the
optically thin MEKAL code, which is a formidable task, in the
following we will assume that, after the plasma becomes optically
thick to radiation at the line frequency, the line luminosity does not
increase if the column density of the gas is increased (see Lazzati et
al. 2002b for discussions on a similar assumptions in reflection
models).

\section{Thermal models for GRB afterglows}

\begin{figure}
\psfig{file=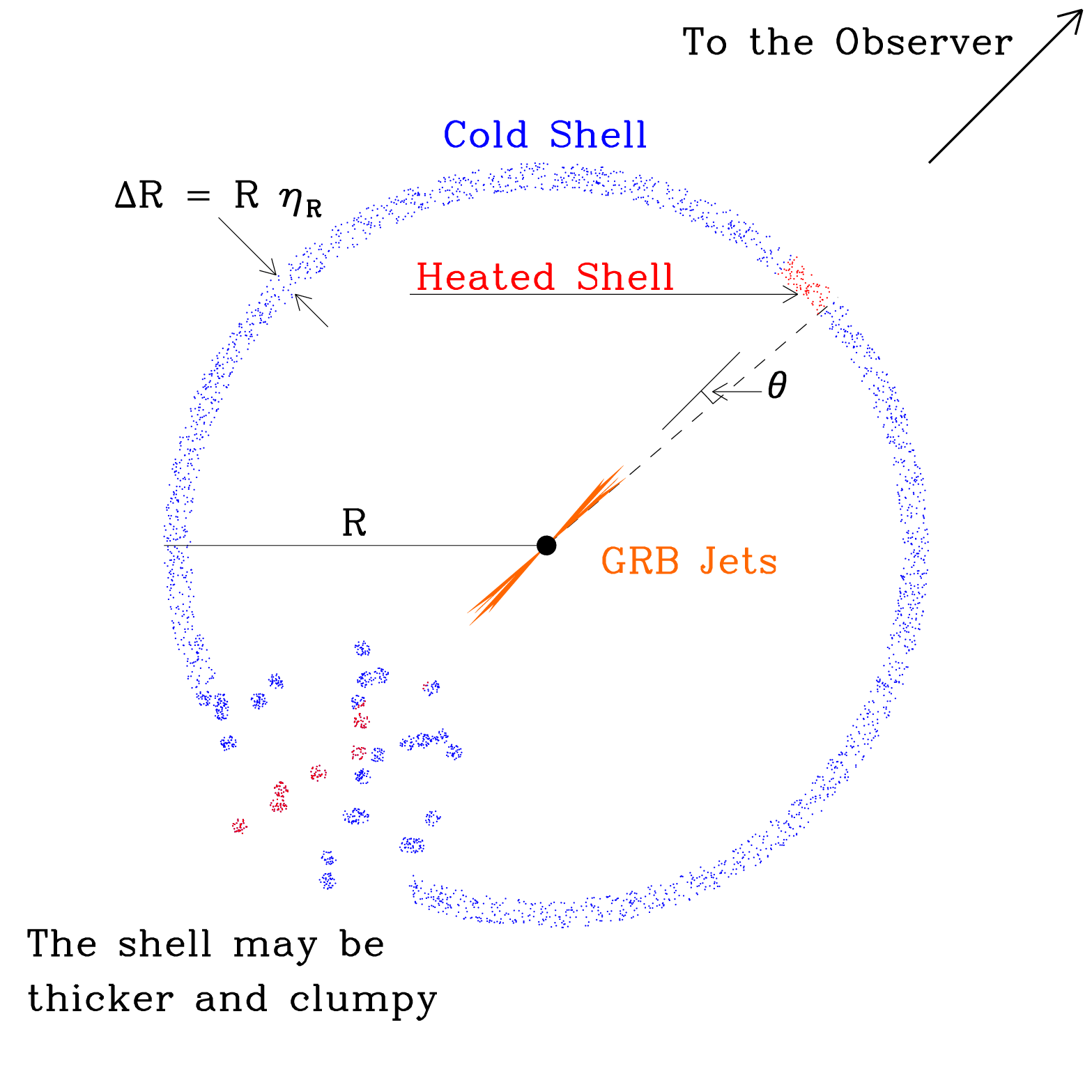,width=0.48\textwidth}
\caption{{Cartoon of the geometrical set-up for the GRB and thermal 
reprocessing material. The GRB is surrounded by a thin shell (or
clumped thicker shell) of material, which is heated by GRB photons
within the fireball solid angle.}
\label{fig:car}}
\end{figure}

Consider a cloud of plasma of electron density $n$ at a temperature
$T$ which covers a solid angle $\Omega=2\pi(1-\cos\theta)$ along the
line of sight to the observer at a distance $R$ from the burst
explosion site (see Fig.~\ref{fig:car}). The luminosity that is
inferred from infinity, taking into account the light travel time
effects, is given by\footnote{Here and in the following we indicate a
quantity $Q$ as $Q=10^x\,Q_x$. We also adopt CGS units.}:
\begin{equation}
L\approx1.7\times10^{-23}\,T_8^{1/2}\,EI\,{{t_{\rm{heat}}+t_{\rm{cool}}}
\over{R/c(1-\cos\theta)+t_{\rm{heat}}+t_{\rm{cool}}}}
\label{eq:lum}
\end{equation}
where $EI$ is the emission integral ($EI=\int_V\,n_en_i\,dV$),
$t_{\rm{cool}}\sim1.2\times10^{15}\,T_8^{1/2}\,n^{-1}$~s is the plasma
cooling time and $t_{\rm{heat}}$ is the heating time, i.e. the time
during which heat (or energy) is supplied to the emitting plasma. It
is assumed that during this heating time the temperature is held
constant, i.e. the plasma is in equilibrium. In the following we will
consider a uniform density plasma and approximate
$n_e\approx{}n_i\approx{}n$. In this case $EI=n^2\,V$, where V is the
volume of the emitting cloud.

In many astrophysically relevant situations (think e.g. to the IGM in
galaxy clusters) $R/c\ll{}t_{\rm{cool}}$ and both the heating and
cooling time scales are irrelevant and Eq.~\ref{eq:lum} simplifies to
the usual free-free equation:
\begin{equation}
L=1.7\times10^{-23}\,T_8^{1/2}\,EI
\label{eq:ff}
\end{equation}
In GRBs this may not be the case. In order to reproduce afterglow
observations in the non-cooling regime, the plasma cloud must satisfy
simultaneously three conditions. First, it must produce the observed
luminosity in the non-cooling regime
\begin{equation}
L=1.7\times10^{-23} \, T_8^{1/2} \, n^2 \, {{4\pi}\over{3}} R^3 \, \eta_R
\, {{\Omega}\over{4\pi}} \sim 10^{46} \, L_{46} \quad {{\rm erg}\over{\rm s}}
\label{eq:c1}
\end{equation}
where the numeric value is typical for the early X-ray afterglow (see
e.g.  GRB~011211, R02) and $\eta_R$ is the volume filling factor of a
possible shell-like or clumpy cloud. Second, it must fulfill the
non-cooling condition:
\begin{equation}
R\,(1-\cos\theta) \le c\,t_{\rm{cool}} = 3.6\times10^{25}\,T_8^{1/2}\,n^{-1}
\label{eq:c2}
\end{equation}
finally, it must require a total amount of energy smaller than the 
total energy of a GRB:
\begin{equation}
E=8.7\times10^{-8} \, T_8 \, n \, R^3 \, \eta_R \, {{\Omega}\over{4\pi}}
\le 10^{52} \, E_{52} \quad {\rm erg}
\label{eq:c3}
\end{equation}

Conditions~\ref{eq:c1} and~\ref{eq:c2} yield the density constraint:
\begin{equation}
n \le 3.7\times10^8 \, T_8^2 \, \eta_R \, L_{46}^{-1} \, {{\Omega}\over{4\pi}}
\label{eq:c4}
\end{equation}
while conditions~\ref{eq:c1} and~\ref{eq:c3} give:
\begin{equation}
n \ge 1.2\times10^9 \, T_8^{1/2} \, L_{46} \, E_{52}^{-1}
\end{equation}
The above conditions cannot clearly be satisfied simultaneously if all
the parameters are taken equal to their fiducial values. Since
$\eta_R$ and $\Omega/4\pi$ are both numbers smaller than unity, a
change in the geometry does not help, making the constraint of
Eq.~\ref{eq:c4} more stringent. Also the temperature cannot be changed
significantly, especially as long as X-ray lines must be taken into
account.

In conclusion, a thermal component that contributes to the early
afterglow of a typical GRB cannot be emitted in the non-cooling
regime. In fact, in order to reach a luminosity large enough without
involving a too large thermal energy, the plasma density must be so
large to make the cooling time very short. As an example, R02 derived
a cooling time of $\sim2$~s for the afterglow of GRB~011211. We cannot
therefore adopt the simple Eq.~\ref{eq:ff} but we must use the more
complete Eq.~\ref{eq:lum}.

We therefore consider in the following two limiting cases:
$R/c(1-\cos\theta)\gg{}t_{\rm{cool}}\gg{}t_{\rm{heat}}$ (flash
heating) and $R/c(1-\cos\theta)\gg{}t_{\rm{heat}}\gg{}t_{\rm{cool}}$
(steady heating). In principle also the case
$t_{\rm{heat}}\gg{}R/c(1-\cos\theta)\gg{}t_{\rm{cool}}$ may be
interesting. It is however difficult to identify a heating source that
can be active for a time comparable to the line emission
time scale. Even if the emitting plasma would be heated by afterglow
photons, the heating time could be at most a few per cent of the line
emission time scale (Lazzati et al. 2002b).

\begin{figure*}
\centerline{\psfig{file=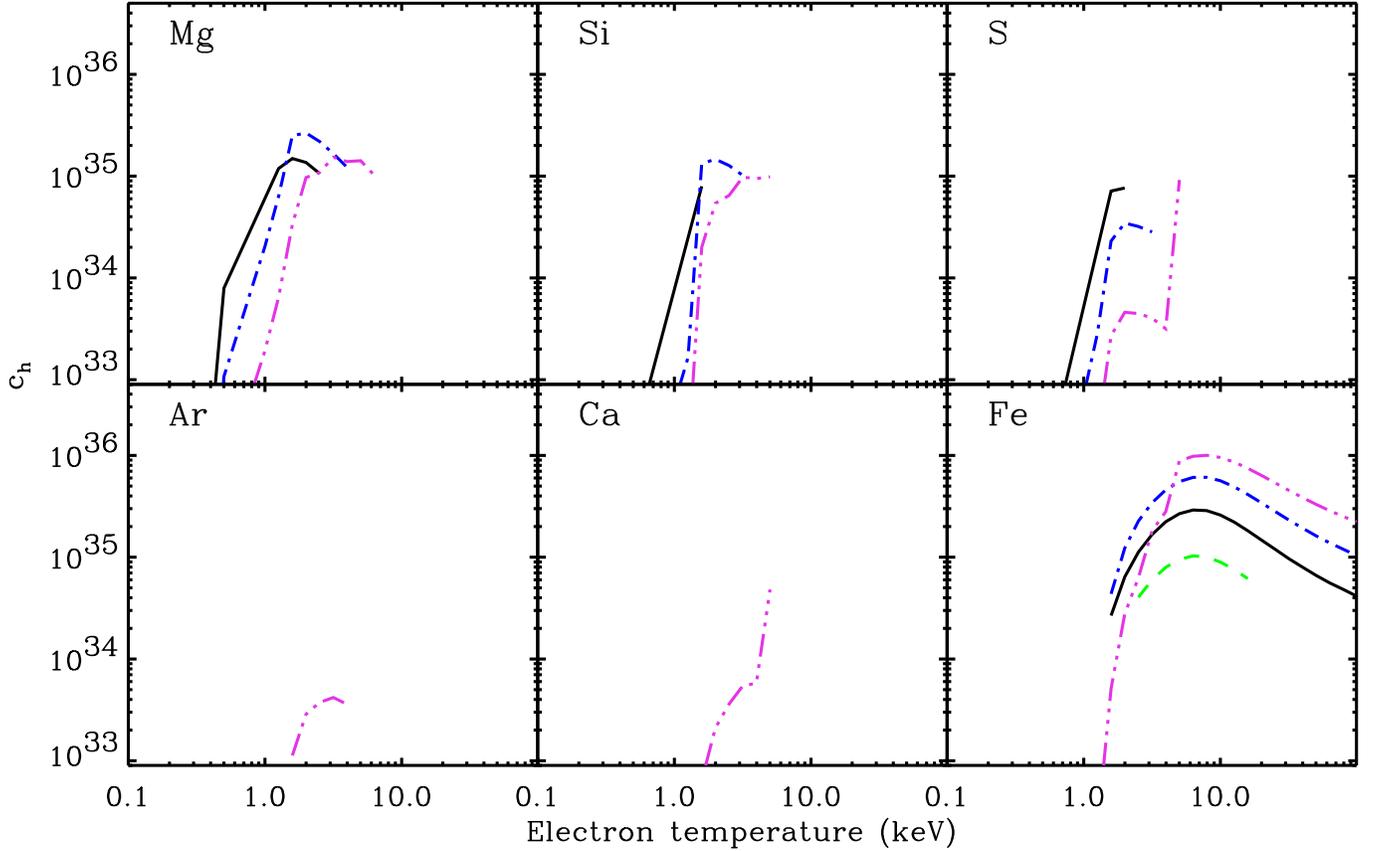,width=\textwidth}}
\caption{{The luminosity coefficient $c_h$ (see text) as a function 
of the plasma temperature and of metallicity for the considered
elements.  As in Fig.~\ref{fig:eta}, the dotted, dashed, solid,
dot-dash and dot-dot-dot-dash lines are relative to 0.1, 0.3, 1, 3 and
10 time solar metallicity, respectively.}
\label{fig:coe}}
\end{figure*}

\subsection{Flash heating}

Consider a portion of a spherical shell of plasma of radius $R$,
temperature $T$ and uniform density $n$ covering a solid angle
$\Omega=1-\cos\theta$ (see, again, Fig.~\ref{fig:car}). Its emission
integral can be written as $EI=n^2\,R^3\,\eta_R\Omega$ where
$\eta_R\equiv\Delta{R}/R$ is the ratio of the thickness of the shell
to its radius\footnote{Note that if a clumpy medium is assumed instead
of a spherical shell, $\eta_R$ is equal to the volume filling factor
of the configuration that maximizes the emitted
luminosity.}. Considering the values $\eta_{\rm{line}}$ defined above
(and shown in Fig.~\ref{fig:eta}), the cloud will produce a line with
luminosity\footnote{The continuum luminosity can be easily derived by
adopting $\eta_{\rm{line}}=1$.}  (see Eq.~\ref{eq:lum}):
\begin{equation}
L_{\rm{line}}=1.9\times10^{27} \eta_{\rm{line}}\,R\,T_8\,\tau_T
\label{eq:lfh}
\end{equation}
where we have used the Thomson opacity
$\tau_T=n\,R\,\eta_R\,\sigma_T$. Eq.~\ref{eq:lfh} coupled with the
efficiencies shown in Fig.~\ref{fig:eta} implies that when we observe
a line with luminosity $L_{\rm{line}}\gsim5\times10^{44}$~erg~s$^{-1}$
(Ghisellini et al. 2002) a thermal model in the flash heating regime
requires a radius larger than $R\gsim2\times10^{17}$~cm. Analogous
results are obtained by imposing a ten times larger continuum
luminosity.  Such a large radius implies that the line luminosity
should remain constant for a time $t=R(1-\cos\theta)/c$, to be
compared with the observed variability time scales
$t_{\rm{var}}\sim2\times10^4$~s observed in e.g. GRB~011211
(R02). This constraint can be satisfied if the heated plasma covers a
very small solid angle, of opening angle $\theta\sim4.5^\circ$, at the
limit but consistent with the smallest opening angles inferred for GRB
jets from afterglow modelling (Frail et al. 2001; Panaitescu \& Kumar
2002). In order to satisfy the cooling condition, however, the plasma
must be dense ($n\gsim10^{11}$~cm$^{-3}$) and therefore confined in a
very thin shell (or small clumps) with $\eta_R\lsim5\times10^{-5}$.
In addition, if we assume that the plasma is due to a spherical shell
surrounding the GRB explosion site which is heated only in the small
polar cap by afterglow photons, the shell mass would turn out to be
$M=4\pi\,R^2\,m_p\,\tau_T/\sigma_T \sim 600\,M_\odot$ for the case
explored above. Both the geometry factor $\eta_R$ and the total mass
required seem to be rather extreme, even though the whole picture is,
in this case, completely self consistent.

\subsection{Steady heating}

Let us now consider a plasma in which a source of heating is active
for a time scale
$t_{\rm{heat}}>1.2\times10^{15}\,T_8/n=t_{\rm{cool}}$. In case of
radiative heating, this time scale may be for example the GRB duration
or even longer, if the afterglow photons can contribute to the
heating.  In steady heating, the line luminosity (continuum luminosity
for $\eta_{\rm{line}}=1$) can be written as:
\begin{equation}
L_{\rm{line}}=7.2\times10^{36}\,\tau_T^2\,T_8^{1/2}\,
\eta_{\rm{line}}{{t_{\rm{heat}}}\over{\eta_R}}
\equiv c_h {{t_{\rm{heat}}}\over{\eta_R}}
\label{eq:lsh}
\end{equation}
where we have defined the parameter $c_h$ in order to emphasize the
ratio of the heating time scale over the geometry parameter.  In
Fig.~\ref{fig:coe} we show the highest possible values of the
coefficient $c_h$ as a function of temperature for five different
values of metallicity (as in Fig.~\ref{fig:eta}). The lines are
plotted only when the equivalent width of the line is predicted to be
larger than 100~eV, corresponding to the unshaded areas in
Fig.~\ref{fig:eta}. 

To compare these results with afterglow data, consider a Fe line with
$L=10^{45}$~erg~s$^{-1}$ (GRB~991216; Piro et al. 2000) or the S line
in GRB~011211 (R02) with luminosity
$L=4\times10^{44}$~erg~s$^{-1}$. In both cases a ratio
$t_{\rm{heat}}/\eta_R\gsim10^9$ is required, by comparison with the
appropriate panel in Fig.~\ref{fig:coe}. A typical heating time can be
considered to be the fireball transit time, which is equivalent to the
total duration of the prompt GRB emission. In both cases such duration
is of the order of $\sim100$~s, leaving us with an extreme requirement
of anisotropy $\eta_R\lsim10^{-7}$. Alternatively, the heating may be
provided radiatively by the absorption of GRB and early afterglow
photons. In that case, the duration of the heat supply may be of
several per cent of the line emission time scale (Lazzati et
al. 2002b). In that case the constraint would be relaxed to the (yet
still challenging) value $\eta_R\lsim10^{-6}$. The most convenient
solution is represented by an (unknown) heating mechanism acting for a
time comparable to the line emission time scale itself. In this case,
the constrain on the geometric factor $\eta_R$ would be similar to the
one derived above for the flash heating case. Such a heating mechanism
is however presently obscure and will have to face the problem of
stability discussed below (\S 4).

In a steady heating scenario, therefore, thermal lines and continua
can dominate the early afterglow emission, only in case of extreme
geometric conditions, in which the emission is produced either in a
sheet like shell or an extremely clumpy medium. Interestingly,
analogous extreme conditions were independently inferred for the
environment of GRB~000210 (Piro et al. 2002) in order to account for
the lack of ionization features in the soft X-ray afterglow
spectrum. Piro et al. (2002, and references therein) argue also that
such conditions may be realized in giant molecular clouds. Stability
considerations (see \S~4), however, show that more extreme conditions
are required in this case. Differently from what derived in the flash
heating condition, in this case the radius of the shell does not have
to be very large, and therefore the total mass required is not huge.

\section{Stability}

In the above section we have derived some geometrical constraints to
the emitting plasma in order to reproduce the observed features and
continua. Since the results envisage particular geometric conditions,
we here analyze their stability. First, since the heating energy is
supposed to come from a central point (the GRB progenitor), one has to
make sure that the emitting material is not accelerated to high
velocities.

Consider a shell of mass $M$ absorbing energy from a relativistic
outflow. If the material is radiative, it acquires a bulk velocity
$v=E/(M\,c)$. A bulk velocity $v\simeq0.1\,c$ was measured in
GRB~011211 (R02) and GRB~991216 (Piro et al. 2000). Requiring that our
optically thin shell is accelerated to a comparable or smaller speed
implies a radius larger than:
\begin{equation}
R\ge{{L\,\sigma_T}\over{4\pi\,m_p\,c^2\,\eta_{\rm{line}}\,\tau_T\,v}}=
3\times10^{13}\,{{L_{\rm{line},45}}\over{\eta_{\rm{line}}\,\tau_T\,v_9}}
\qquad{\rm cm}
\end{equation}
This is not a compelling limit, given the radii discussed above. 

Provided that the emitting medium is not accelerated to relativistic
speeds by the energy input, we also want that the thin emitting shell
(or blobs) do not expand in a time scale smaller than the emission one
(which can be either the heating or cooling time scale).  The shell
(or blobs) was in fact in equilibrium with the ambient medium when it
was cold. Now that its temperature is increased it will tend to expand
under the effect of the increased internal pressure. If it expands at
the speed of sound $c_s$, its density will be sizably modified in a
time scale $t_{\rm{exp}}=R\,\eta_R/c_s$.  It is therefore required
that the expansion time is longer than the emission time scale.

In the case of steady heating we obtain:
\begin{equation}
R>c_s\,{{t_{\rm{heat}}\over{\eta_R}}} = 10^{8}\,T_8^{1/2}\, 
{{t_{\rm{heat}}\over{\eta_R}}} \gsim 10^{17}\,T_8^{1/2}
\end{equation}
Also in the steady heating case, therefore, the radius has to be large
in order to allow for the production of thermal components in the
early GRB X-ray afterglows. For such large radii, however, the density
required to fulfill the steady heating conditions is
$n\gsim10^{15}/t_{\rm{heat}}\gsim10^{12}$, several orders of
magnitudes larger than what inferred by Piro et al. (2002). It seems
therefore that the steady heating case requires more extreme
conditions than the flash heating one (the same distance from the
explosion site and limits on the total mass involved, but larger
densities and smaller filling factors).  Applying the same stability
condition to a flash heating case, we are left with the more relaxed
constraint
\begin{equation}
\tau_T>0.1
\end{equation}
which is therefore not difficult to fulfill.  These stability
considerations suggest therefore that the only viable way to produce a
sizable thermal afterglow component is by heating a small portion of a
massive $\tau_T\sim1$ (either very thin or clumpy) shell of material
located at a relatively large distance from the burst explosion site
(Fig.~\ref{fig:car}). It should however be emphasized that these
stability considerations cannot be applied if the heating is provided
hydrodynamically, in such a way that the source of heating is
providing also the confining pressure. In this case also a steady
heating scenario, with a much smaller shell, may be viable.

\section{Clumpiness of GRB ambient medium}

The above discussion shows that the emission lines detected in several
GRB afterglows can have a thermal origin only if the GRB surroundings
are extremely clumpy, with density and geometric contrasts of order of
a hundred thousand or more. We now discuss the reliability of these
conditions in different GRB progenitor scenario.

In the hypernova scenario, the GRB sets on simultaneously to a
supernova explosion, and the ambient medium is the result of the
interaction of the pre-SN star with its surroundings. In particular,
the nearby ambient will be dominated by the late stages of the mass
ejection history of the star. These stages are known to be unsteady
and clumps or shell-like structures can be envisaged. If the mass
ejection is caused by radiative effects, however, extreme structure
cannot be produced, since the stellar luminosity varies on time scales
comparable to the Kelvin time-scale, which is of the order of hundreds
of years, yielding a thick shell. In the wind environment of SN1998bw,
for example, Li \& Chevalier (1999) found that inhomogeneities up to a
factor of a few were present. A more appealing scenario, in this
perspective, is the supranova model by Vietri and Stella (1998), where
the GRB is supposed to explode several weeks to years after a
supernova. The supernova explosion, having a much smaller time scale,
can generate a more extreme geometry. The clumpiness of SN ejecta has
been investigated by numerical simulations. It is found that
Rayleigh-Taylor instabilities can produce high density clumps with
angular scales $\gsim1^\circ$ (B\"ottcher et al. 2002). Such
structures have $\eta_R\equiv{a}/R\sim10^{-2}$ where $a$ is the clump
radius, much larger than the value required to produce sizable line
emission from collisional excitation. It is however possible, in the
supranova scenario, that the remnant is illuminated by a
super-Eddington relativistic wind (Vietri \& Stella 1998, Konigl \&
Granot 2002) in the time span between the SN and GRB explosions. The
interaction of this wind with the ejected SN shell may increase the
inhomogeneities originally present in the shell (Guetta \& Granot
2003; Lazzati \& Rees in preparation).

\section{Discussion}

In at least three XMM-Newton spectra of X-ray GRB afterglows, the
presence of a thermal component has been claimed (R02; Watson et
al. 2002). We have studied the geometrical conditions that the thermal
material must satisfy in order to contribute significantly to the
early X-ray afterglow. We first concluded that the time scale during
which the emission is observed must be set by the light crossing time
of the emitting region. This imply that standard free-free equations
that relate the emission integral to the luminosity cannot be used and
a more general equation (Eq.~\ref{eq:lum}) has to be adopted. The
implications of this are of great importance. In fact, the use of the
standard formalism (Eq.~\ref{eq:ff}) to compute the EI led R02 to
underestimate the emission integral for GRB~011211 by a factor
$\sim10^4$, and conclude that thermal emission from a shell with
$R\sim10^{15}$~cm can be a self consistent solution. We showed that,
with the correct treatment of light-crossing effects, this is not the
case. We then studied the possibility of (i) flash heating and (ii)
steady heating. In both cases, we conclude that the thermal material
must be extremely clumped in order to contribute significantly to a
typical X-ray afterglow.  In the steady heating case, the thermal
material is continuously heated for a time longer than the cooling
one, reaching a stationary state (albeit for a time smaller than the
light-crossing one). We showed that in this case the material has to
be so extremely clumped that the clumps are dissolved by the increased
temperature in a very short time scale, so that an additional source
of confinement must be envisaged. In the case of flash heating, a self
consistent solution can be found, requiring a less extreme (even if
yet compelling) clumping. In this case, the thermal material must be
located far from the source and therefore, in order to preserve the
fast time variability, only a small portion of it has to be
heated. Assuming that the same density is spread all over the GRB
explosion site, we derive that the total mass involved is large ($\sim
600\,M_\odot$), larger than even a type-II supernova remnant. A
moderate asymmetry of the remnant could however ameliorate this
requirement. It is interesting to note that, even though such
clumpy and dense environments seem extreme, they were already inferred
independently to account for the lack of ionization features in
GRB~000210 (Piro et al. 2002) and to explain the haigh energy emission
in GRB~940217 (Katz 1994).

To date there is not yet a compelling evidence that there are indeed
thermal components in the early X-ray afterglow of GRBs. It is however
intriguing then in three out of four afterglows observed by XMM-Newton
a thermal model yields to a better fit than an absorbed power-law
(R02, Watson et al. 2002a). In the case of GRB~011211 (R02) is
particularly difficult to explain the absence of an iron or nickel
line, while reflection models seem to give a better explanation to the
observed line ratios (Lazzati et al. 2002b).

Nonetheless, should a thermal component be confirmed in the X-ray
afterglow of at least a sub-fraction of GRBs, it would be difficult to
avoid the conclusion that (i) the GRB explosion site is surrounded by
a massive shell of matter, similar to a 1--2 years old supernova
remnant (as expected in some versions of the Supranova model, Vietri
\& Stella 1998); (ii) that this remnant is extremely clumped, 
suggesting that some precursor activity has been taking place. Whether
such a clumpiness can help making the prompt emission in an external
shock scenario (Dermer et al. 2000) heavily depends on the presence of
clumping at smaller scales, a test that cannot be done with thermal
emission from the clumps, since would contribute in any case a
negligible flux to the early afterglow emission.

\acknowledgements{I am grateful to G. Ghisellini, P. Kumar, M. J. Rees 
and F. Tavecchio for useful comments and discussions. I acknowledge
financial support from the PPARC.}

\end{document}